\documentclass[sip,biber]{now-journal} 
\usepackage[utf8]{inputenc}
\usepackage{amssymb}
\usepackage{float}

\usepackage{comment}
\usepackage{graphicx}
\usepackage{subcaption}
\usepackage{algorithm}
\usepackage{algpseudocode}
\usepackage{amsmath,amssymb}
\usepackage{xcolor} 

\title{FLRSP: Privacy-Preserving Federated Learning Using Randomly Selected Model Parameters}

\author[1]{Sawada,Hiroto}
\author[1]{Imaizumi,Shoko}
\author[2]{Kiya,Hitoshi}

\affil[1]{Chiba University, Japan}
\affil[2]{Tokyo Metropolitan University, Japan}

\issuevolumeyear{2025}
\issuevolumenumber{xx}
\issueelocation{e} 
\articletype{Original Paper} 

\articledatabox{Received xx xxxxx 2026; Revised xx xxxxx 2026\\[2pt]
ISSN 2048-7703; DOI 10.1561/116.xxxxxxxx\\
\copyright\ 2026 H. Sawada, S. Imaizumi and H. Kiya}

\OAstatement{This is an Open Access article, distributed under the terms of the Creative Commons Attribution licence (\url{http://creativecommons.org/licenses/by-nc/4.0/}), which permits unrestricted re-use, distribution, and reproduction in any medium, for non-commercial use, provided the original work is properly cited.}


\creditline*{Corresponding author: 
Hitoshi Kiya, kiya@tmu.ac.jp. This study was partially supported by JSPS KAKENHI (Grant Number JP25K07750) and JST CREST (Grant NumberJPMJCR20D3).}

\begin{document}

\begin{abstract}
 In this paper, we propose a method for privacy-preserving federated learning that uses randomly selected model parameters to update global models. High-quality deep neural networks (DNN) models require a huge amount of training data in general, but model training raises privacy concerns when dealing with sensitive or personal information. Federated learning is a distributed machine learning framework in which multiple clients and a server train a model collaboratively.
 However, if the shared updates are compromised, an attacker may reconstruct the original training data.
 In addition, previous methods for improving robustness generally reduce the accuracy.
 To overcome these issues, in our method called federated learning using randomly selected model parameters (FLRSP), model parameters computed in each local server are randomly selected and shared to update a global model in a central server. In experiments, image classification tasks were carried out on the ResNet34 architecture and the Vision Transformer (ViT) under the use of Federated Stochastic Gradient Descent (FedSGD) and Federated Averaging (FedAvg), and the results demonstrated our method's effectiveness in terms of image classification accuracy and robustness against state-of-the-art attacks compared with previous methods.
\end{abstract}

\section{Introduction}

Training a high-quality deep neural network (DNN) model requires a huge amount of training data, but model training raises privacy concerns because training data generally includes sensitive or personal information \cite{cite_0,cite_kiya_1,cite_kiya_2,cite_kiya_3,cite_kiya_4}.
To collect a huge amount of training data, in centralized machine learning, each client prepares information on local data and sends it to a central
server for model training. Federated learning (FL) \cite{cite_1} is a learning approach that allows multiple clients to collaboratively train a shared model while keeping
their data decentralized and private. In FL, clients participate in the training
process using their local data without sharing their raw data.

However, several studies have pointed out that some attack methods infer training data from the updated information shared by clients \cite{cite_2,cite_3,cite_4,cite_5,cite_6,cite_AT, cite_addAT_11,cite_addAT_12,cite_addAT_13,cite_addAT_14,cite_addAT_15}.
Accordingly, several methods for
improving the security of FL, (hereafter "security enhancement methods") have been studied to address this. For differential privacy \cite{cite_11,cite_12,cite_13,cite_14,cite_15,cite_addDP_1,cite_addDP_2,cite_addDP_3,cite_addDP_4,cite_addDP_5,cite_addDP_7,cite_addDP_8,cite_addDP_9,cite_addDP_10}, specific noise is added to updated information to make it more resistant to attacks; however, this degrades model accuracy. 

Additionally, Lu et al. proposed a security enhancement method that is robust to the gradient leakage attack called Attention PRIvacy Leakage (APRIL) \cite{cite_6}.
An effective security enhancement
method using perceptual encryption was also proposed \cite{cite_20}. 
However, both of these methods are only applicable to Vision Transformer (ViT)-based models. 
That is, they cannot be applied to convolutional neural networks (CNNs).

In addition, the initial concept of using random binary weights was introduced in ViT and confirmed that the method is robust against APRIL \cite{cite_19}.
It was shown in \cite{cite_19} that using randomly selected gradients makes image restoration attacks more difficult by not sharing a portion of the updated information, while the random selection mechanism is expected to update all parameters eventually. In \cite{cite_19}, information updated at each client is randomly selected at each epoch; therefore, all parameters of the global model are expected to be updated eventually as training proceeds.
However, the discussion in \cite{cite_19} is limited to ViT and FedSGD.

To address this problem, we expand the initial concept presented in \cite{cite_19} and confirm its effectiveness. We propose federated learning using randomly selected model parameters (FLRSP) as well in which each client randomly selects model parameters before sharing them with the server.
The basic concept proposed in \cite{cite_19} is expanded and generalized in the present paper. The main additions include the use of FedAvg in addition to FedSGD, the use of ResNet in addition to ViT, a comparison with differential privacy protection methods, and an evaluation of resilience against image restoration based on adversarial optimization attacks.
In experiments, image classification tasks are carried out on the ViT and ResNet34, and the results demonstrate the effectiveness of FLRSP in terms of model accuracy and attack resistance.

The contributions of this paper are as follows.
\begin{quote}
\begin{enumerate}
  \renewcommand{\labelenumi}{(\alph{enumi})}
  \setlength{\leftskip}{20pt} 

  \item 
  We propose a novel FL method to defend against attackers by selecting randomly selected parameters of the global model computed at each local server.

  \item We verify that the proposed method outperforms state-of-the-art methods in terms of robustness against attacks and classification accuracy on image classification tasks.
\end{enumerate}
\end{quote}
The rest of this paper is structured as follows. Section \ref{section_2} presents related studies on FL, attack methods, and security enhanced methods.
Section \ref{section_3} gives an overview and includes the details of the proposed method, FLRSP. Experiments verifying the effectiveness of the method, including classification accuracy, robustness against attacks, and discussion are presented in Section \ref{section_4}, and Section \ref{section_5} concludes this paper.

\section{Related Work}
\label{section_2}

\subsection{Federated Learning}

\begin{figure}[tb]
\begin{center}
\includegraphics[width=\linewidth]{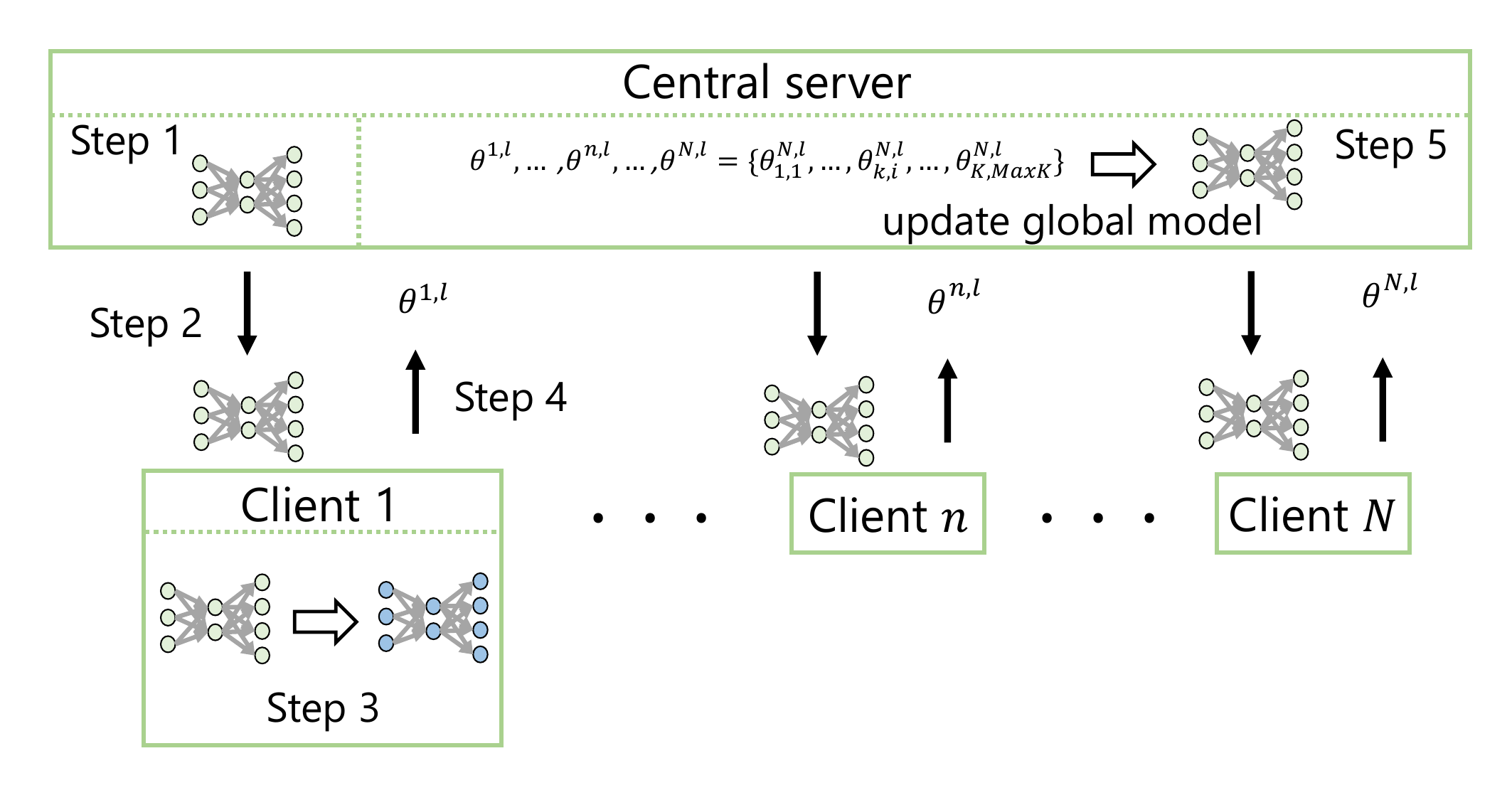}
\caption{Overview of FL under FedSGD.}
\label{fig1_FL}
\end{center}
\end{figure}

FL is a distributed machine learning method that is executed by multiple clients holding datasets and a central server providing learning models, as shown in Fig.\ref{fig1_FL} \cite{cite_1}.
The procedure for standard FL is as follows.

\begin{quote}
\renewcommand{\labelenumi}{\bf Step\ \theenumi:}
\begin{enumerate}
\setlength{\leftskip}{20pt} 
{\item \noindent
A central server initializes a global model.

\item \noindent
The central server distributes its current global model to each client.

\item \noindent
Each client locally trains the global model with its own training data.

\item \noindent
Each client sends information on the updated model  to the server.

\item \noindent
The server aggregates the updated information received from clients and updates the global model.

}
\end{enumerate}
\end{quote}
The above steps 2 to 5 are repeated for model training.
Each client only sends updated information such as gradients, weights, and bias values to the server, so the dataset is not shared.

Here, we focus on Federated Stochastic Gradient Descent (FedSGD) and Federated Averaging (FedAvg) as key algorithms used in FL for updating the global model. In FedSGD, each client sends the gradients of their model back to the central server for each batch training. The server aggregates gradients received from clients and uses them to update the global model.
In FedAvg, which is an extension of FedSGD, each client sends the weights and bias values of its model to the central server for each epoch of training.
The server aggregates these parameters and computes their arithmetic mean to set as the parameters of the global model.

\subsection{Image Restoration Attacks}

\begin{figure}[tb]
\begin{center}
\includegraphics[width=\linewidth]{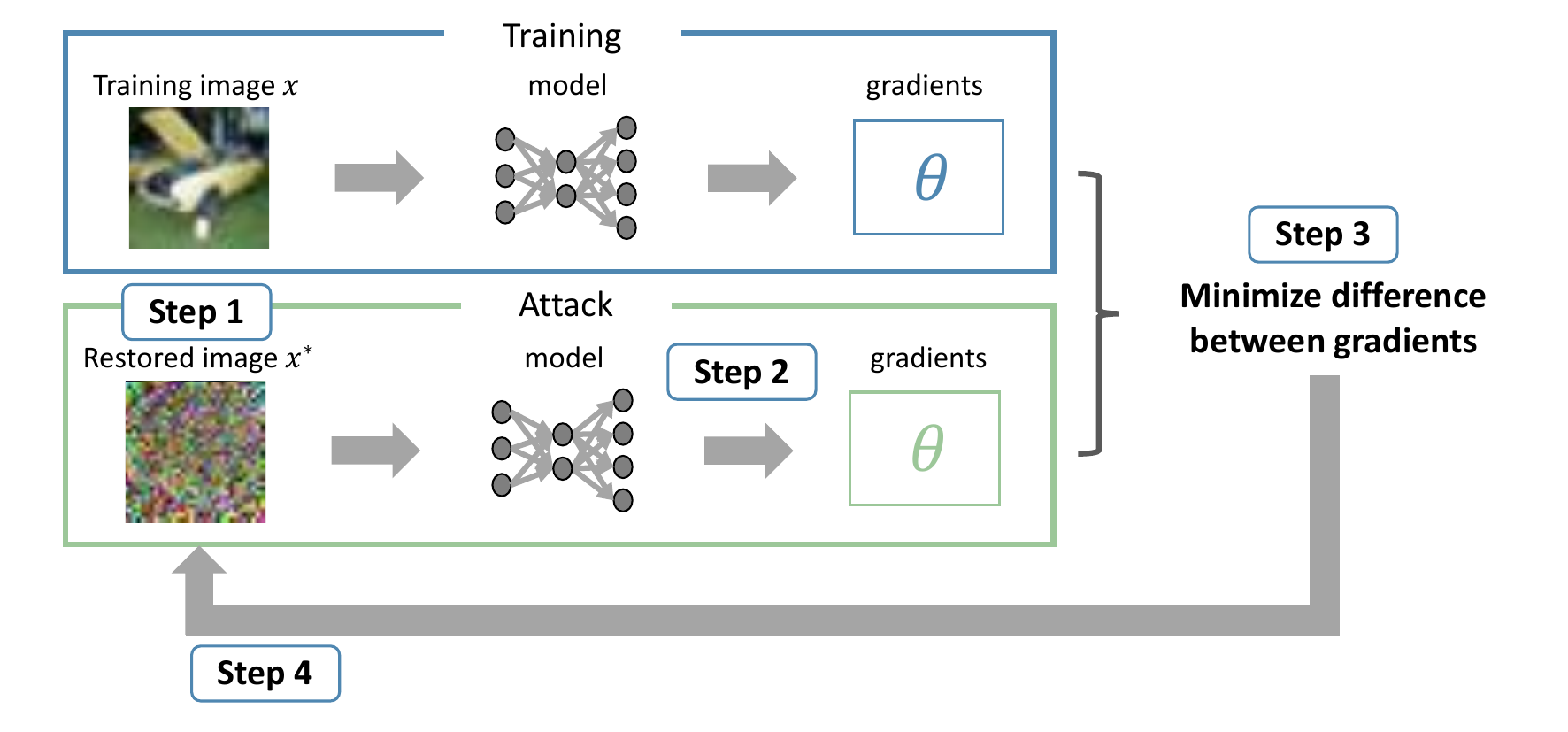}
\caption{Strategy based on adversarial attacks in FL.}
\label{fig6_AT}
\Description{figure description ABCABC}
\end{center}
\end{figure}

In FL, clients can share updated information instead of training data for learning.
However, there are attacks that aim to restore training data from the updated information \cite{cite_2,cite_3,cite_4,cite_5,cite_6,cite_member}.
For example, there are reconstruction attacks that enable image restoration across multiple batches \cite{cite_2}.
In addition, membership inference can verify whether a specific image is used for training \cite{cite_member}.
Furthermore, a property inference attack can infer attributes unrelated to the target task \cite{cite_pia}.
Among these, we focused on Attention PRIvacy Leakage (APRIL) \cite{cite_6, cite_APRIL_2, cite_APRIL_3} and optimization strategies \cite{cite_AT, cite_3, cite_CNNAT_3}, referred to as adversarial optimization attacks, as state-of-the-art attacks, which can restore a single image from updated information with high accuracy.

The aim of an attacker is to restore visual information from updated model information shared by clients. We assume that the attacker is able to use updated model information shared and the security enhanced algorithm by the clients including probability $R$, under this condition, our method is evaluated through simulations using two types of attacks that are effective against ViT and CNNs, respectively.

APRIL \cite{cite_6} is an image restoration attack specifically designed for ViT. It can infer input images by using the updated information from a target client.
APRIL consists of two steps.
In the first step, a sequence of embedded patches in ViT $z_0 \in \mathbb{R}^{(P_s + 1) \times P^2 \cdot C}$ is calculated from
\begin{equation}
\frac{\partial l}{\partial z_{0}} z_{0}^{T} = q1^{T} \cdot \frac{\partial l}{\partial q1} + v1^{T} \cdot \frac{\partial l}{\partial v1} + k1^{T} \cdot \frac{\partial l}{\partial k1},
\label{eq_APRIL_z0_1}
\end{equation}
\begin{equation}
\frac{\partial l}{\partial z_{0}} = \frac{\partial l}{\partial E_{pos}}.
\label{eq_APRIL_z0_2}
\end{equation}
Here, $P_s$ is the number of patches, $P$ is the patch size, $C$ is the number of channels, $ \frac{\partial l}{\partial \sharp} $ is an output of the loss function for $\sharp$, $E_{pos} \in \mathbb{R}^{(P_s + 1) \times D}$ is a tensor of the positional embedding layer, and $D$ is the length of the embedding vector.
Additionally, $q1$, $k1$, and $v1 \in \mathbb{R}^{(P_s + 1) \times D_h}$ are the first tensors of the transformer encoder layer, referred to as query, key, and value, respectively, and $D_h$ is  the vector length $D$ divided by the number of self-attention in ViT.
In the second step, a restored image $x*$ is obtained by
\begin{equation}
x* = E_{patch} \times (z_{0} - E_{pos})^{T},
\label{eq_APRIL_x}
\end{equation}
where $E_{patch} \in \mathbb{R}^{(P^2 \cdot C) \times D}$ is a matrix of the first linear layer. Note that APRIL is carried out using a closed-form equation as in Eq.\eqref{eq_APRIL_x} to restore images.

Adversarial optimization attacks are known to be effective for CNNs \cite{cite_AT}.
Here, each client trains the model iteratively, and attackers restore the training data by using the updated information at any time during the learning process.
The attack procedure is as follows (see Fig.\ref{fig6_AT}).

\begin{quote}
\renewcommand{\labelenumi}{\bf Step\ \theenumi:}
\begin{enumerate}
\setlength{\leftskip}{20pt} 
{\item \noindent
Generate a random image $x^{*}$.

\item \noindent
Input the generated image to the model and compute the gradients of the image.

\item \noindent
Calculate the difference between the computed gradients and gradients shared by the client.

\item \noindent
Update the generated image as a restored image by using the difference.
} 
\end{enumerate}
\end{quote}
The above steps 2 through 4 are repeated a defined number of times, and the restored image is expected to closely match the training data used by clients.
Here, since bias operation corresponds to a simple addition to input data, the gradients of the bias can be regarded as those of training images as
\begin{equation}
\frac{dl}{dx_j} = \frac{dl}{db_j}.
\label{eq_attack_bias}
\end{equation}
Here, $\frac{dl}{dx_j}$ denotes the gradient of the $j$-th element of the training image $x$, and $b_j$ means the gradient of the $j$-th bias.
Therefore, a restored image $x^{*}$ can be iteratively updated by minimizing the difference between the gradients of the bias.
The difference is given by

\begin{equation}
 \frac{<\bigtriangledown_{\theta} L_{\theta} \left( x, y\right), \bigtriangledown_{\theta} L_{\theta} \left( x^{*}, y\right)>}{||\bigtriangledown_{\theta} L_{\theta} \left( x, y\right)||\ ||\bigtriangledown_{\theta} L_{\theta} \left( x^{*}, y\right)||}.
\label{eq_attack}
\end{equation}
Here, $\bigtriangledown_{\theta} L_{\theta} \left( x, y\right)$ is a set of gradients for an input image $x$ with a label $y$.
This attack can restore training images by using the cosine similarity between gradients shown in Eq.\eqref{eq_attack}.

\subsection{Security Enhancement Methods}
Unauthorized use of data and models prepared for machine learning need to be protected from various attacks and accidents \cite{cite_kiya_1,cite_kiya_2,cite_kiya_3,cite_kiya_4}, so methods for enhancing the security of FL have also been actively pursued.
In this paper, we mainly focus on the following two enhancement methods, which are effective against APRIL and widely used across various domains for enhancing security, respectively.

First, the fixed-position method is a security enhancement approach for ViT \cite{cite_6}.
This method sets all updated information in the positional embedding layer of ViT to zero.
If the ViT's position embedding layer is pre-trained, omitting updates has minimal impact on model performance.
Therefore, the fixed-position method can provide attack resistance without decreasing accuracy.
However, when training a model without pre-training, this method degrades model performance.

Second, differential privacy is known as a representative example of security enhancement methods for FL \cite{cite_11,cite_12,cite_13,cite_14,cite_15,cite_addDP_1,cite_addDP_2,cite_addDP_3,cite_addDP_4,cite_addDP_5,cite_addDP_7,cite_addDP_8,cite_addDP_9,cite_addDP_10}.
This method ensures security by adding a specific noise to the updated information.
In this paper, we use Gaussian differential privacy to determine the noise as in \cite{add_DP_125}.
A noisy gradient $\hat{\theta}_{k, i}^{n, l}$ is calculated by the following equation.
\begin{equation}
\begin{aligned}
\hat{\theta}_{k, i}^{n, l} &= \theta_{k, i}^{n, l} + N\left( 0, S_f^{2}\sigma^{2} \right),\\
\sigma &\geq \frac{\sqrt{2ln\left(1.25/\delta \right)}}{\epsilon}.
\end{aligned}
\end{equation}

Here, $\sigma$ is a noise scale, $S_f$ is the sensitivity of models, $\epsilon$ is the privacy intensity, and $\delta$ is the probability at which differential privacy is not satisfied.
Differential privacy aims to protect updated information with arbitrary strength.
Security can be enhanced as $\epsilon$ and $\delta$ decrease.
However, there is a trade-off between privacy protection and model accuracy.
To address this, we propose a novel privacy protection method that preserves the model accuracy.

\section{Federated Learning with Randomly Selected Parameter}
\label{section_3}
\begin{figure}[t]
\begin{center}
\begin{minipage}[b]{0.9\linewidth}
\includegraphics[width=\linewidth]{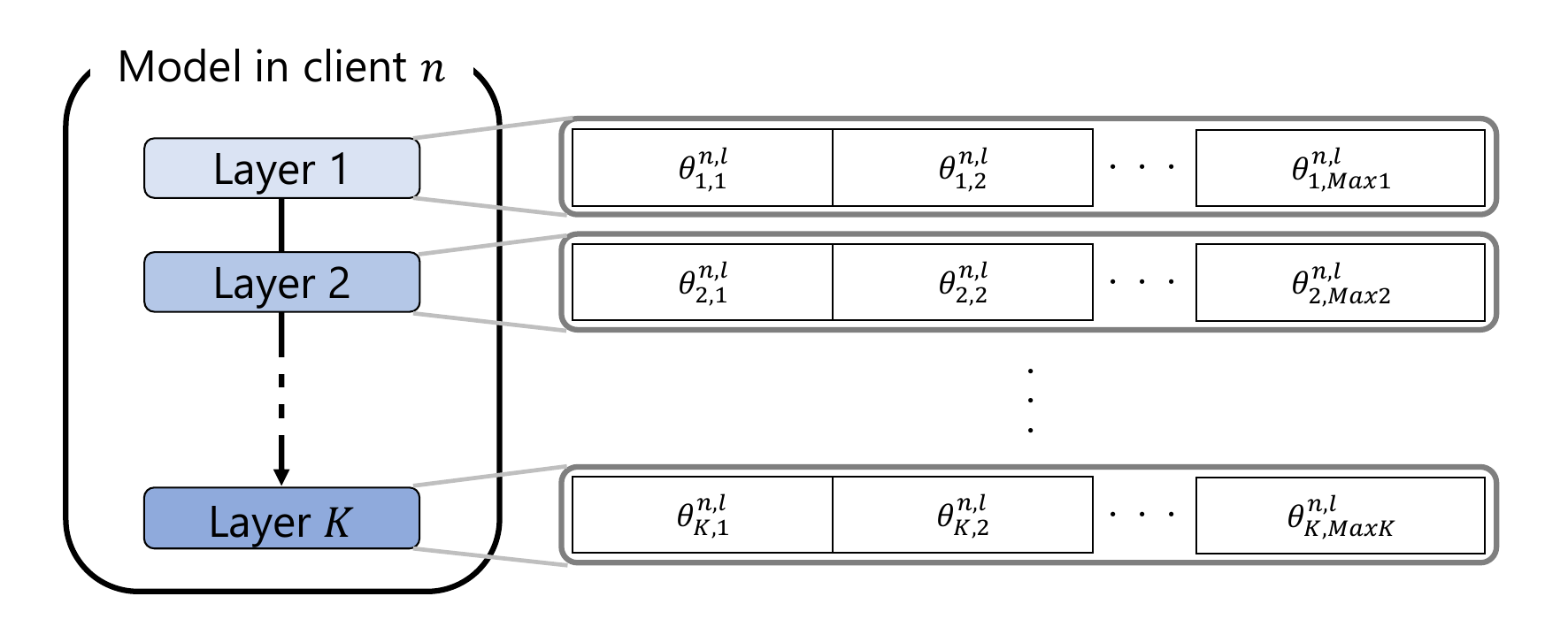}
\subcaption{Gradients \texorpdfstring{$\theta_{k, i}^{n, l}$}\ used in standard FedSGD}
\label{fig:3_Normal}
\end{minipage}
\\
\begin{minipage}[b]{0.9\linewidth}
\includegraphics[width=\linewidth]{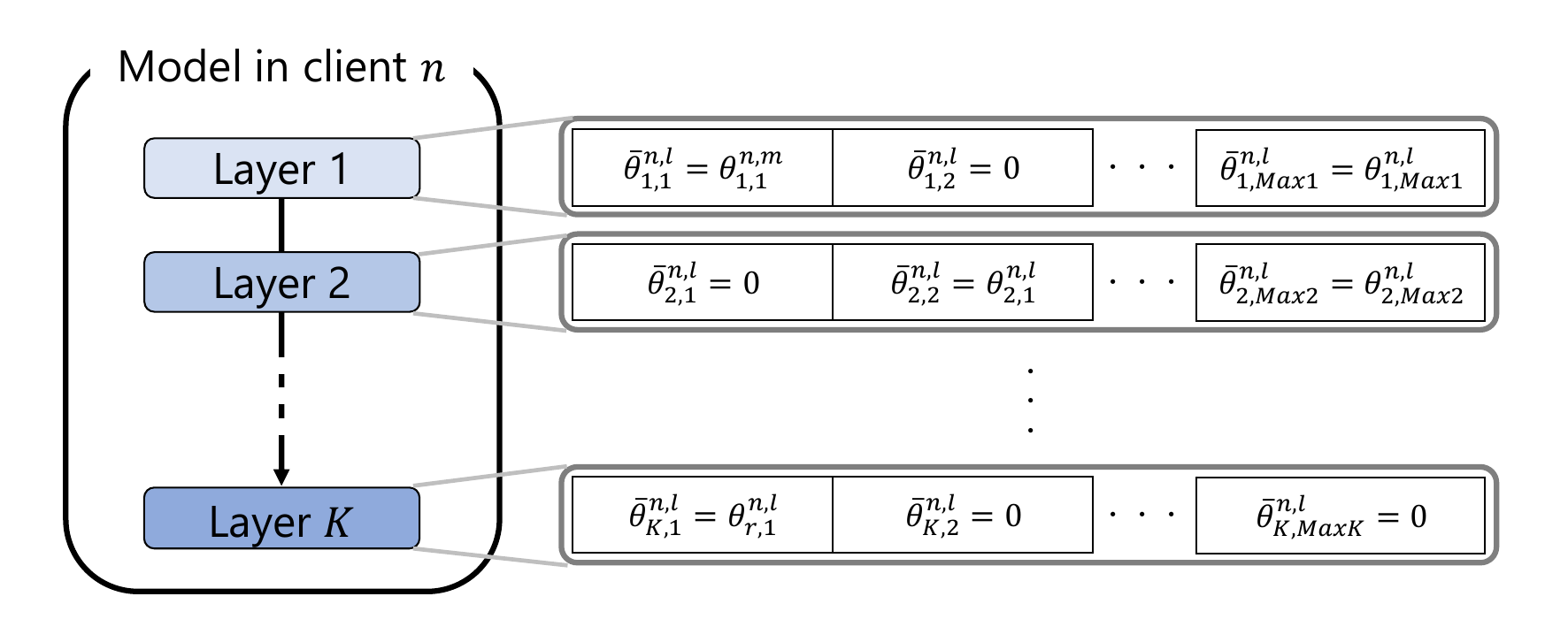}
\subcaption{Randomly selected
 gradients \texorpdfstring{$\overline{\theta}_{k, i}^{n, l}$}\  used in our method}
\label{fig:4_Random}
\end{minipage}
\end{center}
\centering
\caption{Randomly selected
 model parameters for FedSGD.}
\Description{figure description ABCABC}
\label{fig:Random_binary_weights}
\end{figure}
In our proposed method, called federated learning with randomly selected parameters (FLRSP), model parameters calculated by each client using their own data are randomly selected before being aggregated on a central server. The method aims to maintain model accuracy while also providing robustness against attacks.

\subsection{Overview of Proposed Method}

The procedure of the proposed method is given below.

\begin{quote}
\renewcommand{\labelenumi}{\bf Step\ \theenumi:}
\begin{enumerate}
\setlength{\leftskip}{20pt} 
{\item \noindent
A central server initializes a global model.

\item \noindent
The central server distributes its current global model to each client.

\item \noindent
Each client locally trains the global model with its own training data.

\item \noindent
Each client randomly selects the parameters of the updated model.

\item \noindent
Each client sends information on the selected parameters to the server.

\item \noindent
The server aggregates the parameters received from clients and updates the global model.
}
\end{enumerate}
\end{quote}
The above procedure is repeated for training.
Model weights and bias values in the global model are updated each time.
The method can be applied to both FedSGD and FedAvg as follows.

\subsection{FLRSP for FedSGD}

\begin{algorithm}[htbp]
\caption{FedSGD with FLRSP. Number of clients $N$,
learning rate $\eta$,
The probability of zero occurrence $R$,
number of rounds/batches $l$,
maximum number of rounds/batches $L$,
loss $\mathcal{L}$,
and parameters of global model in the $l$-th
batch ${w}^{l}$.}
\label{alg}
\begin{algorithmic}
\Statex
\State \textbf{Server executes:}
\State Initialize global model ${w}^{0}$
\State Send ${w}^{0}$ to all clients $n=1,\dots,N$
\For{$l = 0,\dots,L-1$} 
    \State receive gradients computed by clients,
    \ForAll{clients \textbf{in parallel}}
    \State $\bar{{\theta}}^{\,n,l}
    \gets \textsc{ClientUpdate}(n,l,{w}^{l},R),$
    \State $\bar{{\theta}}^{\,n,l}
    = \{\bar{\theta}_{1,1}^{n,l}, \dots,
    \bar{\theta}_{K,\mathrm{MaxK}}^{n,l}\}$
    \EndFor
    \State Aggregate gradients and update global model:
    \[
        {w}^{l+1} \leftarrow {w}^{l} - \eta\  \frac{1}{\sum_{n=1}^{N}B^{n,l}}\sum_{n=1}^{N}\bar{{\theta}}_{}^{n, l}
    \]
    \begin{center}
    \State ${w}^{l+1} \leftarrow {w}^{l}$, if $\sum_{n=1}^{N}B^{n,l} = 0$
    \end{center}
    \State Send updated model ${w}^{l+1}$ back to all clients
\EndFor
\State \Return ${w}^{L}$
\Statex
\Function{ClientUpdate}{$n,l,{w},R$} \Comment{Run on client $n$}
    \State Sample a mini-batch $({x}^{n,l},{y}^{n,l})$
    \State Compute gradient of loss:
    \State $f({x}^{n,l};{w})$ is the output of global model $w$
    \[
        {\theta}^{\,n,l} \leftarrow \nabla_{{w}}
        \mathcal{L}\!\left(f({x}^{n,l};{w}),{y}^{n,l}\right), {\theta}^{\,n,l}
    = \{\theta_{1,1}^{n,l}, \dots,
    \theta_{K,\mathrm{MaxK}}^{n,l}\}
    \]
    \State Element of a binary random sequence \State ${B}^{n,l}=\{{B}_{1,1}^{n,l}, \dots,
    {B}_{K,\mathrm{MaxK}}^{n,l}\}, \ {B}_{k,i}^{n,l} \in \{0, 1\}$:
    \[
        B^{n,l} \sim \mathrm{Bernoulli}(R)
    \]
    \State Randomly zero out gradients (client-side):
    \[
        \bar{{\theta}}^{\,n,l} \leftarrow {B}^{n,l}\odot {\theta}^{\,n,l}
    \]
    \State \Return $\bar{{\theta}}^{\,n,l}$ to server
\EndFunction
\end{algorithmic}
\end{algorithm}

First, we propose an FL method that carries out FedSGD with randomly selected model parameters.
In FedSGD, the server updates the global model for each batch learning using the gradients of model parameters shared from each client.
As in step 3, each client locally trains the model distributed from the central server with its own training data.
In this learning process, each client obtains gradients $\theta_{k, i}^{n, l}$ in the $l$-th batch.
Here, $\theta_{k, i}^{n, l}$ is the $i$-th gradient in the $k$-th layer calculated with an optimization function by the client $n$.

As indicated in step 4, each client randomly selects gradients of model parameters as
\begin{equation}
 \overline{\theta}_{k, i}^{n, l} = B_{k, i}^{n, l} \times \theta_{k, i}^{n, l},
 \ (i=1, 2, ..., Maxk),
 \ (k=1, 2, ..., K),
\label{eq_proposed_multi}
\end{equation}
where $B_{k, i}^{n, l} \in \{ 0, 1 \} $ is the element of a binary random sequence used for selecting parameters, and $Maxk$ is the number of parameters in the $k$-th layer.
In standard FL, each client generally computes the gradients of all parameters and sends them to the server as in Fig.\ref{fig:Random_binary_weights}\subref{fig:3_Normal}. In contrast, in FLRSP, some of the gradients are replaced with zero values as in Fig \ref{fig:Random_binary_weights}\subref{fig:4_Random}.
Accordingly, several terms of $\frac{\partial l}{\partial q1*}$, $\frac{\partial l}{\partial k1*}$, $\frac{\partial l}{\partial v1*}$, $\frac{\partial l}{\partial E_{pos}*}$, and $\frac{\partial l}{\partial b_{j}*}$, in Eqs. \eqref{eq_APRIL_z0_1}, \eqref{eq_APRIL_z0_2}, and \eqref{eq_attack_bias} become zeros, thereby preventing the equations required for the image restoration attacks from being satisfied.

In FLRSP, each client can independently determine random binary sequences. The probability $R$ of zero occurrence in a random sequence is decided, and random sequences are then generated. High values of $R$ increase the probability that the model will not be updated properly but can enhance robustness.

Next, we describe how the global model is updated with gradients received from clients on the central server as in step 6. In FLRSP, the following equation for gradient integration is carried out,

\begin{equation}
  w_{k, i}^{l+1}=
  \begin{cases}
    w_{k, i}^{l} - \eta \frac{\sum_{n=1}^N  \overline{\theta}_{k, i}^{n, l}}{\sum_{n=1}^{N} B_{k, i}^{n, l}} & \text{if $0 < \sum_{n=1}^{N} B_{k, i}^{n, l} \leq N$} \\
    w_{k, i}^{l}                 & \text{if $\sum_{n=1}^{N} B_{k, i}^{n, l} = 0$}, \\
  \end{cases}
\label{eq_FedSGD_t_v}
\end{equation}
where $w_{k, i}^{l}$ is the $i$-th model parameter of the $k$-th layer in the $l$-th batch, $\eta$ is a learning rate, and $N$ is the number of clients.
This equation reveals that the proposed method only refers to non-zero gradients when determining the average of the gradients. Accordingly, the global model can be updated even if many parameters are replaced with zero values. 
Note that Eq.\eqref{eq_FedSGD_t_v} is reduced to standard FL when $R=0$ is selected.
The overall procedure of the proposed method for FedSGD is summarized in Algorithm \ref{alg}.

\subsection{FLRSP for FedAvg}

Next, we describe our method for FedAvg.
In FedSGD, each client updates the global model for each batch, but updating the global model is carried out with the mean of parameters for each epoch in FedAvg.
In this update, the $i$-th parameter of the $k$-th layer obtained by the client $n$ in the $m$-th epoch is denoted as $W_{k, i}^{n, m}$.

Here, parameters are selected during step 4 by the proposed method as
\begin{equation}
\overline{W}_{k, i}^{n, m} = B_{k, i}^{n, m} \times W_{k, i} ^{n, m}.
\label{eq_FedAvg_}
\end{equation}
The proposed method achieves privacy protection by using these selected parameters.
As described in step 5, the selected parameters are sent to the server.
In addition, the parameters of the global model $W_{k, i}^{m}$ are updated by using parameters $\overline{W}_{k, i}^{n, m}$ received from clients as given by:
\begin{equation}
  W_{k, i}^{m+1}=
  \begin{cases}
    \frac{\sum_{n=1}^N  \overline{W}_{k, i}^{n, m}}{\sum_{n=1}^{N} B_{k, i}^{n, m}} & \text{if $0 < \sum_{n=1}^{N} B_{k, i}^{n, m} \leq N$} \\
    W_{k, i}^{m}                 & \text{if $\sum_{n=1}^{N} B_{k, i}^{n, m} = 0$}. \\
  \end{cases}
\label{eq_FedAvg_t}
\end{equation}

In FedAvg, the above parameter update is repeated for each epoch.

\subsection{Selection of Probability \texorpdfstring{$R$}\ }
\label{sec_34}

The target requirements of the proposed method are as follows.

\begin{quote}
\renewcommand 
{\labelenumi}{\bfseries (\alph{enumi}):}
\begin{enumerate}
\setlength{\leftskip}{20pt} 
{\item \noindent
Accuracy comparable to that of models trained with the standard FL.

\item \noindent
Robustness against various attacks from global model creators, other clients, and external third parties.


}
\end{enumerate}
\end{quote}
As shown later in experiments, selecting a high probability $R$ enhances robustness against attacks but also increases the impact on learning.
Therefore, it is necessary to set an appropriate R based on the application of federated learning.

A suitable value of R generally depends on the task, the number of clients, the degree of non-i.i.d, and the assumed attack methods. Anticipated attacks and requirements are determined according to the model to be used, and an effective probability $R$ against the attacks and requirements are experimentally selected in advance. 
For example, one criterion for determining R is to keep the Structural Similarity Index Measure (SSIM) of the reconstructed image below 0.5 \cite{cite_0224_SSIM}.
This threshold value was discussed in \cite{cite_0224_thre}, which concluded that private information in the original image becomes visually difficult to identify.
Furthermore, our experimental results indicated that the random seed has little dependence on SSIM. Therefore, the value of $R$ can be determined in relatively few experiments.

\subsection{Convergence Analysis}
In addition, we discuss the convergence property of FLRSP in FedSGD.
The following condition describes the convergence of FedSGD under stadard FL~\cite{cite_0224_syusoku}:
\begin{equation}
\begin{split}
\mathbb{E}\left[ \left\| w^{l+1} - \ddot{w}^{l+1} \right\| \right]
&\le (1+\eta L_c)^l \, \mathbb{E}\left[ \left\| w^{1} - \ddot{w}^{1} \right\| \right] \\
& + \eta \sum_{i=1}^{l} (1+\eta L_c)^{l-i}
\mathbb{E}\left[ \left\| \nabla F\left(w^{i}, S_i\right) - \nabla F\left(w^{i}, N_{a}\right) \right\| \right],
\end{split}
\label{eq_1_syusoku}
\end{equation}
where $w^{l}$ indicates model parameters in the l-th batch, $\ddot{w}^{l}$ is the ideal parameters in the $l$-th batch, $L_c$ is a Lipschitz constant, $S_i$ is a set of the clients that participate in the i-th batch, $N_a$ is a set of all clients, $\nabla F\left(w^{l}, S_l\right)$ is the average gradient of the loss for the clients collected in the $l$-th batch and $\nabla F\left(w^{i}, N_a\right)$ is the average gradient of the loss for all clients in the $l$-th batch.
This inequality represents the maximum discrepancy in model updates between standard learning and federated learning.

When FLRSP is applied, a learning rate $\eta'$ is given as $\eta' = \eta(1 - R^N)$.
Therefore, the convergence condition is rewritten as
\begin{equation}
\begin{split}
\mathbb{E}\left[ \left\| w^{l+1} - \ddot{w}^{l+1} \right\| \right]
&\le (1+\eta' L_c)^l \, \mathbb{E}\left[ \left\| w^{1} - \ddot{w}^{1} \right\| \right] \\
& + \eta' \sum_{i=1}^{l} (1+\eta' L_c)^{l-i}
\mathbb{E}\left[ \left\| \nabla F\left(w^{i}, S_i\right) - \nabla F\left(w^{i}, N_{a}\right) \right\| \right].
\end{split}
\label{eq_2_syusoku}
\end{equation}
The above mathematically shows that the learning process converges even when the proposed method is applied.
The left-hand side of the inequality indicates the difference between the model trained without FL and that trained with FLRSP, and the right-hand side denotes a non-negative value. For standard FL, $\eta'$ is replaced with $\eta$ in (12). Accordingly, if standard FL has stable convergence for updating the global model, the FLRSP setting also provides stable convergence, since the only difference between FLRSP and standard FL is the learning rate and the rate takes a positive value less than or equal to 1. 

\section{Experimental Results}
\label{section_4}
We conducted image classification experiments to verify the effectiveness of the proposed method in terms of attack resistance and model accuracy.

\subsection{Setup}

\begin{table}[t]
    \centering
    \caption{Experimental conditions}
    \label{tab1}
    \begin{tabular}{|c|c|c|}
        \hline
        \# of clients $N$ & \multicolumn{2}{|c|}{5} \\ \hline
        Model & ViT & ResNet \\ \hline
        Patch size & 16 & -- \\ \hline
        Learning rate $\eta$ & 0.00001 & 0.0001 \\ \hline
        Dataset & \multicolumn{2}{|c|}{CIFAR-10} \\ \hline
        Batch size & \multicolumn{2}{|c|}{32} \\ \hline
        Occurrence probability of zeros $R$ & \multicolumn{2}{|c|}{0.2, 0.5, and 0.8} \\ \hline
        Privacy intensity $\epsilon$ & \multicolumn{2}{|c|}{1, 2, and 4} \\ \hline
        Failure probability of differential privacy $\delta$ & \multicolumn{2}{|c|}{0.5} \\ \hline
    \end{tabular}
\end{table}

Image classification tasks were carried out on the CIFAR-10 dataset consisting of 50,000 training images and 10,000 test images.
As shown in Table \ref{tab1}, the number of clients was set to $N = 5$.
Each client had 10,000 training images without duplication. 
The model architectures used were vit\_small\_patch16\_224 and ResNet34.
The learning rate $\eta$ and patch size for ViT were 0.00001 and 16, respectively, while the learning rate $\eta$ for ResNet was 0.0001.
Each client trained their local model using all of the training images with a batch size of 32.
Probability $R$ in the proposed method was set to 0.2, 0.5, and 0.8.
In differential privacy, the privacy intensity $\epsilon$ was set to 1, 2, and 4.
Furthermore, the probability $\delta$ at which the differential privacy guarantee may not hold was set to 0.5 as in \cite{cite_14}.

\subsection{Image Classification Accuracy}
First, the effectiveness of FLRSP is verified in terms of classification accuracy for 10 epochs.
The accuracy of image classification $Acc$ is given by
\begin{equation}
  Acc = \frac{T}{C},
\label{eq_FedSGD_t}
\end{equation}
where $T$ is the number of correctly classified images and $C$ is the number of test images.


\subsubsection{Impact on ViT Performance}
\begin{figure}[t]
  \centering
    \begin{subfigure}[t]{0.48\linewidth}
    \centering
    \includegraphics[width=6cm]{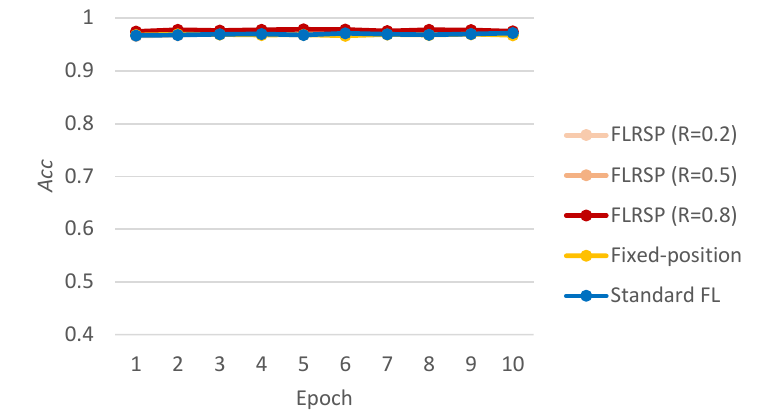}
    \caption{Fine-tuning pre-trained models}
    \label{train_v1}
  \end{subfigure}
  \centering
    \begin{subfigure}[t]{0.48\linewidth}
    \centering
    \includegraphics[width=6cm]{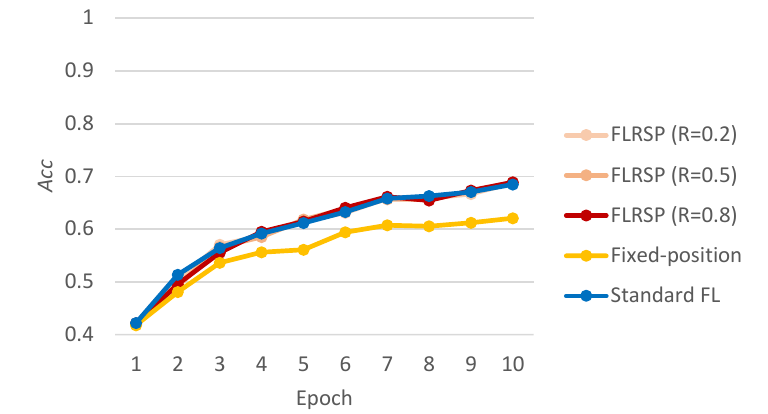}
    \caption{Training models from scratch}
    \label{train_v2}
  \end{subfigure}
\caption{Accuracy of ViT under FedSGD.}
\label{fig5_g_1}
\end{figure}

We first compare our method with the fixed-position method \cite{cite_6}, which is a strong security enhancement method proposed for ViT.
Fig.\ref{fig5_g_1} depicts accuracy results at each epoch for fine-tuning a pre-trained ViT on ImageNet \cite{cite_17} (see Fig.\ref{fig5_g_1}\subref{train_v1}) and for the training models from scratch approach (see Fig.\ref{fig5_g_1}\subref{train_v2}), respectively, where standard FL is FL without any security enhancement.

From Fig.\ref{fig5_g_1}\subref{train_v1}, it is evident that both our method and the fixed-position method maintained accuracy comparable to that of standard FL. In contrast, as shown in Fig.\ref{fig5_g_1}\subref{train_v2}, the accuracy of the fixed-position method degraded, although our method demonstrated almost the same accuracy as that of baseline models. When using models from scratch, initial parameters are typically unsuitable for target tasks, so all layers of ViT should be updated. However, the fixed-position method does not update the positional embedding layer. Accordingly, the accuracy of the fixed-position method is degraded when using models from scratch, but our method can update all layers using randomly selected model parameters.


\subsubsection{Impact on ResNet Performance}
\begin{figure}[t]
  \centering
    \begin{subfigure}[t]{0.48\linewidth}
    \centering
    \includegraphics[width=6cm]{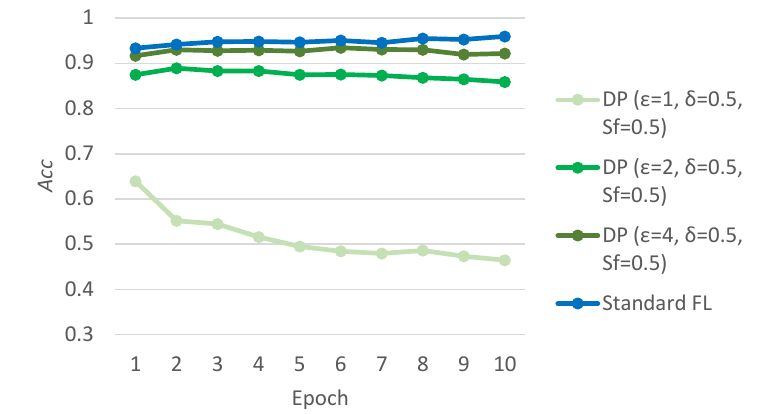}
    \caption{DP}
    \label{train_2}
  \end{subfigure}
  \centering
  \begin{subfigure}[t]{0.48\linewidth}
    \centering
    \includegraphics[width=6cm]{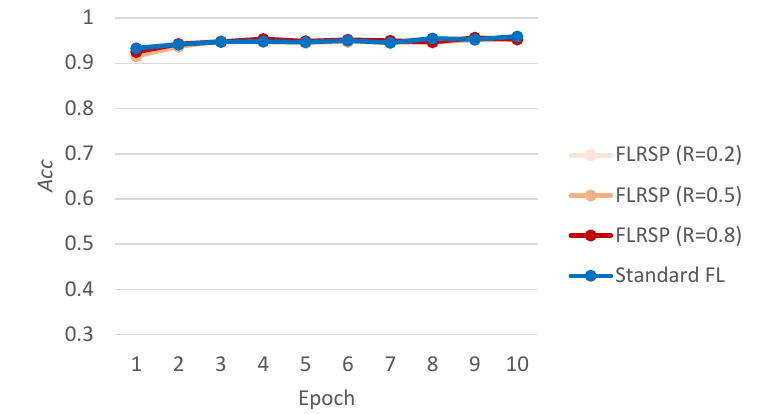}
    \caption{FLRSP}
    \label{train_1}
  \end{subfigure}
\caption{Accuracy of ResNet34 under FedSGD.}
\label{fig5_g_2}
\end{figure}

\begin{figure}[t]
  \centering
  \centering
    \begin{subfigure}[t]{0.48\linewidth}
    \centering
    \includegraphics[width=6cm]{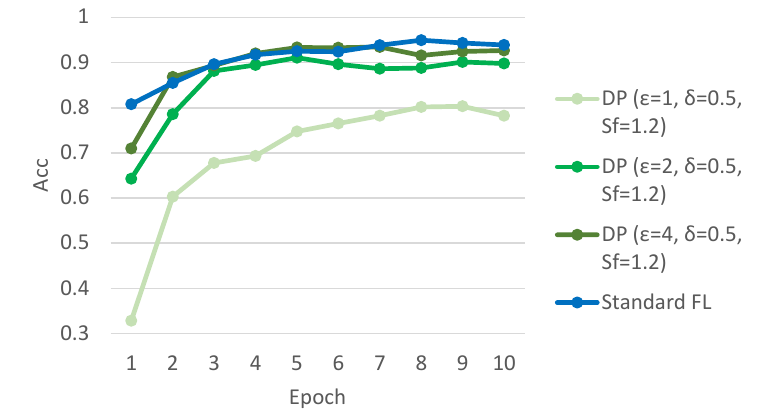}
    \caption{DP}
    \label{train_3}
  \end{subfigure}
  \centering
    \begin{subfigure}[t]{0.48\linewidth}
    \centering
    \includegraphics[width=6cm]{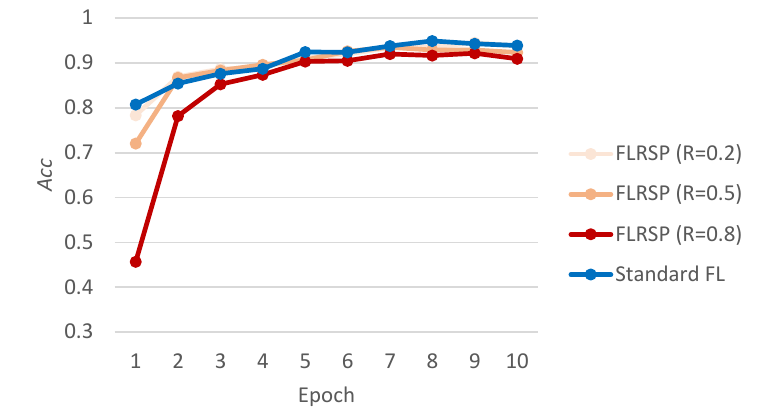}
    \caption{FLRSP}
    \label{train_4}
  \end{subfigure}\\
\caption{Accuracy of ResNet34 under FedAvg.}
\label{fig5_g}
\end{figure}

\begin{figure}[t]
  \centering
  \centering
    \begin{subfigure}[t]{0.48\linewidth}
    \centering
    \includegraphics[width=6cm]{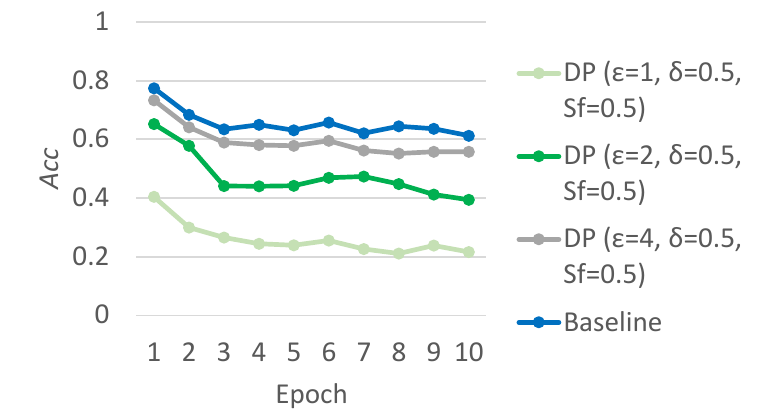}
    \caption{DP}
    \label{train_5}
  \end{subfigure}
  \centering
    \begin{subfigure}[t]{0.48\linewidth}
    \centering
    \includegraphics[width=6cm]{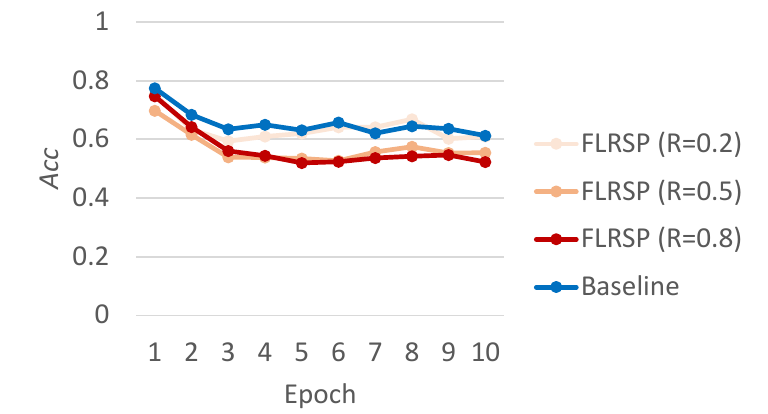}
    \caption{FLRSP}
    \label{train_6}
  \end{subfigure}\\
\caption{Accuracy of ResNet34 under FedSGD with non-i.i.d. dataset.}
\label{g_0224_non}
\end{figure}

Next, our method was compared with the framework of differential privacy (DP) on ResNet34 pre-trained with ImageNet \cite{cite_online2}. Fig.\ref{fig5_g_2}\subref{train_2} is the result of ResNet models enhanced with DP under FedSGD.
As the figure shows, the accuracy degradation caused by noise was improved by selecting a large value of $\epsilon$.
However, as will be described later, selecting such a large $\epsilon$ reduced robustness against attacks.
Therefore, a small value of $\epsilon$ should be chosen for image restoration attacks when using DP.

In contrast, FLRSP maintained high accuracy under the use of various $R$ values as in Fig.\ref{fig5_g_2}\subref{train_1}.
We determined that the proposed method did not degrade classification accuracy compared to standard FL when $R = 0.2, 0.5,$ and $0.8$.
Therefore, our method can satisfy requirement (a) in \ref{sec_34}.

Fig.\ref{fig5_g}\subref{train_3} shows the results of FedAvg with DP applied to ResNet, in which DP also reduced accuracy as $\epsilon$ decreased.
Although classification accuracy hardly decreased with $\epsilon=4$, it became difficult to prevent attacks.

The results of our method with FedAvg are presented in Fig.\ref{fig5_g}\subref{train_4}.
FLRSP was found to exhibit low accuracy during the initial learning phase.
In particular, it decreased significantly for $R=0.8$.
In FedSGD, gradients per batch are shared for updating the global model, while in FedAvg, parameters per epoch are shared. Therefore, compared to FedAvg, FedSGD generally results in smaller update values, so the magnitude of $R$ appears to have little impact in the FeDSGD case. For example, in Figures 5 and 6, we confirmed that FedSGD shared gradients ranging from 0.01 to 0.0001 between the client and server, while FedAvg shared parameters ranging from 10 to 0.1 between the client and server. This difference reduces the impact of $R$ in FedSGD compared to FedAvg.
FLRSP also contributes to strengthening the resistance of FedAvg, because the basic principle of FedAvg for strengthening resistance is the same as that of FedSGD.
However, the models ultimately achieved a similar level of accuracy to standard FL.
These results demonstrate that FLRSP can achieve high accuracy close to that of standard FL by continuing to learn if an appropriate $R$-value is selected.

\subsubsection{Impact of non-i.i.d datasets}
In realistic federated or distributed scenarios, data heterogeneity (non-i.i.d. distributions) is common, so we conducted additional experiments where the Dirichlet distribution with hyperparameter $a=0.1$, which is used as a standard metric to evaluate how well it can tolerate bias in realistic data, was used for evaluation (see Fig. \ref{g_0224_non}). 
From Figs. \ref{fig5_g_2} and \ref{g_0224_non}, we see that the accuracy of FLRSP was affected by the non-i.i.d distribution similar to DP, but FLRSP maintained still higher accuracy than that of DP. Relatively high classification accuracy was also maintained at low $R$ values. In addition, for non-i.i.d. datasets, adjustments are generally made to improve model performance using the learning rate and loss. The proposed method is also expected to improve accuracy through such adjustments, but the specific details of these adjustments will be a subject for future research.

Moreover, since the attacks use only the updated information of the target training images, privacy leakage characteristics do not differ under heterogeneous distributions.


\subsection{Robustness of Security Enhancement Methods}

Next, FLRSP was evaluated in terms of robustness against image restoration attacks. We utilized APRIL \cite{cite_6} proposed for ViT and an optimization strategy based on adversarial attacks \cite{cite_AT} targeting CNN models as state-of-the-art attack methods.
We simulated each attack using 15 images.

In this paper, we mainly focus on FedSGD when discussing attack resistance. The reason is that FedSGD is known to have more stable convergence and less influence from non-i.i.d datasets compared to FedAvg, but its attack resistance is lower \cite{0224_AT_Avg}. If attack resistance can be sufficiently strengthened in FedSGD, FLRSP would also contribute to strengthening the resistance of FedAvg, because the basic principle for strengthening resistance is the same as that of FedSGD.

We evaluated the robustness against attacks using the SSIM \cite{cite_0224_SSIM} between the training image and the restored image.
The SSIM between two images $x_1$ and $x_2$ is defined as
\begin{equation}
\mathrm{SSIM}(x_1, x_2) = \frac{(2 \mu_{x_1} \mu_{x_2} + C_1)(2 \sigma_{x_{1}x_{2}} + C_2)}{( \mu_{x_{1}}^2 + \mu_{x_{2}}^2 + C_1)(\sigma_{x_{1}}^2 + \sigma_{x_{2}}^2 + C_2)},
\end{equation}
\label{eq_SSIM}
where $\mu_{{x}_{1}}$ denotes the mean values of $x_1$, similarly, $\mu_{{x}_{2}}$ indicates the mean values of $x_2$, $\sigma_{x_{1}}^2$ denotes the variances, $C_1$ and $C_2$ are small constants for preventing division by zero, and $\sigma_{x_{1} x_{2}}$ indicates the covariance.

\subsubsection{Resistance against Attention PRIvacy Leakage}

\begin{figure}[t]
\begin{center}
\begin{minipage}[t]{0.3\linewidth}
\begin{center}
\includegraphics[width=2.2cm]{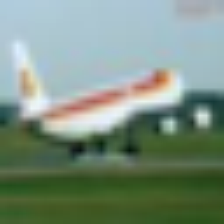}
\end{center}
\subcaption{Training image}
\label{fig:4_1}
\end{minipage}
\begin{minipage}[t]{0.3\linewidth}
\begin{center}
\includegraphics[width=2.2cm]{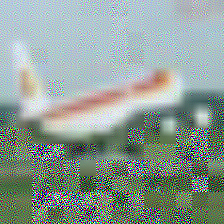}
\end{center}
\subcaption[width=2cm]{Standard FL}\textit{}
\label{fig:4_2}
\end{minipage}
\begin{minipage}[t]{0.3\linewidth}
\begin{center}
\includegraphics[width=2.2cm]{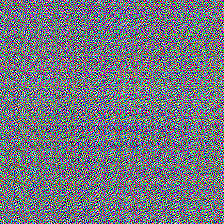}
\end{center}
\subcaption[width=2cm]{Fixed-position}
\label{fig:4_3}
\end{minipage}
\end{center}
\begin{center}
\begin{subfigure}{\textwidth}
    \centering

    \begin{minipage}{0.3\textwidth}
      \centering
      \includegraphics[width=2.2cm]{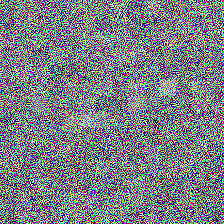}
      \caption*{R=0.2}
    \end{minipage}
    \begin{minipage}{0.3\textwidth}
      \centering
      \includegraphics[width=2.2cm]{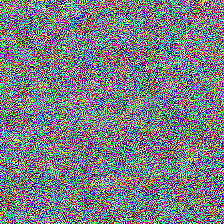}
      \caption*{R=0.5}
    \end{minipage}
    \begin{minipage}{0.3\textwidth}
      \centering
      \includegraphics[width=2.2cm]{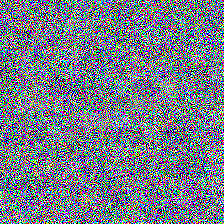}
      \caption*{R=0.8}
    \end{minipage}
\end{subfigure}
\subcaption{FLRSP}
\label{img_APRIL_FLRSP}
\end{center}
\caption{An example of images restored by APRIL.}
\label{fig:4_1_6}
\end{figure}

First, we verified the resistance to attack against APRIL, which is well-known as a strong attack proposed for the FL of ViT. Fig.\ref{fig:4_1_6} illustrates an example of images restored with APRIL.
Fig.\ref{fig:4_1_6}\subref{fig:4_1} is a training image and \ref{fig:4_1_6}\subref{fig:4_2} shows a restored image for standard FL.
As shown in Fig.\ref{fig:4_1_6}\subref{fig:4_2}, APRIL fully disclosed the visual information of the training image for standard FL.
On the other hand, as shown in Fig.\ref{fig:4_1_6}\subref{fig:4_3}, the fixed-position method was robust against restoration by APRIL. Similarly, Fig.\ref{fig:4_1_6}\subref{img_APRIL_FLRSP} shows that FLRSP provided effective safety enhancements against APRIL.
It was also confirmed that FLRSP had similar trends for other images.
The proposed method replaces a subset of the elements in the gradients with zeros. Consequently, Eqs. \eqref{eq_APRIL_z0_1} and \eqref{eq_APRIL_z0_2} do not hold, making it difficult to reconstruct images.
Therefore, we determined that FLRSP possesses robustness equivalent to that of the fixed-position method against APRIL even when using $R=0.2$.

\subsubsection{Resistance against Adversarial Optimization Attacks}
\setcounter{subfigure}{0}
\begin{figure}[htbp]
  \centering
  \begin{subfigure}{\textwidth}
    \centering
    \begin{minipage}{0.2\textwidth}
      \centering
      \includegraphics[width=\linewidth]{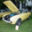}
    \end{minipage}
    \begin{minipage}{0.2\textwidth}
      \centering
      \includegraphics[width=\linewidth]{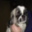}
    \end{minipage}
    \begin{minipage}{0.2\textwidth}
      \centering
      \includegraphics[width=\linewidth]{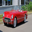}
    \end{minipage}
    \begin{minipage}{0.2\textwidth}
      \centering
      \includegraphics[width=\linewidth]{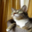}
    \end{minipage}
    \caption{Original training images}
    \label{fig3_a}
  \end{subfigure}
  \vspace{1em}
  \begin{subfigure}{\textwidth}
    \centering
    \begin{minipage}{0.2\textwidth}
      \centering
      \includegraphics[width=\linewidth]{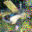}
      \caption*{SSIM = 0.789}
    \end{minipage}
    \begin{minipage}{0.2\textwidth}
      \centering
      \includegraphics[width=\linewidth]{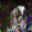}
      \caption*{SSIM = 0.746}
    \end{minipage}
    \begin{minipage}{0.2\textwidth}
      \centering
      \includegraphics[width=\linewidth]{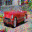}
      \caption*{SSIM = 0.868}
    \end{minipage}
    \begin{minipage}{0.2\textwidth}
      \centering
      \includegraphics[width=\linewidth]{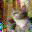}
      \caption*{SSIM = 0.830}
    \end{minipage}
    \caption{Restored images \texorpdfstring{$(Acc = 0.960)$}}
    \label{fig3_b}
  \end{subfigure}
  \caption{Images restored by adversarial optimization attacks under standard FL}
  \label{fig3}
\end{figure}
\captionsetup[subfigure]{labelformat=simple}
\begin{figure}[htbp]
  \centering
  \begin{subfigure}{\textwidth}
    \centering
    \begin{minipage}{0.2\textwidth}
      \centering
      \includegraphics[width=\linewidth]{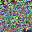}
      \caption*{SSIM = 0.087}
    \end{minipage}
    \begin{minipage}{0.2\textwidth}
      \centering
      \includegraphics[width=\linewidth]{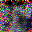}
      \caption*{SSIM = 0.101}
    \end{minipage}
    \begin{minipage}{0.2\textwidth}
      \centering
      \includegraphics[width=\linewidth]{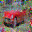}
      \caption*{SSIM = 0.677}
    \end{minipage}
    \begin{minipage}{0.2\textwidth}
      \centering
      \includegraphics[width=\linewidth]{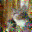}
      \caption*{SSIM = 0.772}
    \end{minipage}
    \caption{\texorpdfstring{$\epsilon=1,\delta=0.5,$}\ \  and \texorpdfstring{$S_f=0.5\ (Acc=0.492)$}}
    \label{fig_dp_a}
  \end{subfigure}
  \vspace{1em}
  \begin{subfigure}{\textwidth}
    \centering
    \begin{minipage}{0.2\textwidth}
      \centering
      \includegraphics[width=\linewidth]{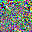}
      \caption*{SSIM = 0.061}
    \end{minipage}
    \begin{minipage}{0.2\textwidth}
      \centering
      \includegraphics[width=\linewidth]{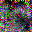}
      \caption*{SSIM = 0.110}
    \end{minipage}
    \begin{minipage}{0.2\textwidth}
      \centering
      \includegraphics[width=\linewidth]{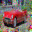}
      \caption*{SSIM = 0.826}
    \end{minipage}
    \begin{minipage}{0.2\textwidth}
      \centering
      \includegraphics[width=\linewidth]{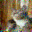}
      \caption*{SSIM = 0.844}
    \end{minipage}
    \caption{\texorpdfstring{$\epsilon=4,\delta=0.5,$}\ \ and \texorpdfstring{$S_f=0.5\ (Acc=0.926)$}}
    \label{fig_dp_b}
  \end{subfigure}

  \caption{Images restored by adversarial optimization attacks under DP.}
  \label{fig_dp_all}
\end{figure}
\begin{figure}[t]
  \centering
  \begin{subfigure}{\textwidth}
    \centering
    \begin{minipage}{0.2\textwidth}
      \centering
      \includegraphics[width=\linewidth]{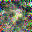}
      \caption*{SSIM = 0.237}
    \end{minipage}
    \begin{minipage}{0.2\textwidth}
      \centering
      \includegraphics[width=\linewidth]{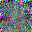}
      \caption*{SSIM = 0.122}
    \end{minipage}
    \begin{minipage}{0.2\textwidth}
      \centering
      \includegraphics[width=\linewidth]{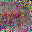}
      \caption*{SSIM = 0.160}
    \end{minipage}
    \begin{minipage}{0.2\textwidth}
      \centering
      \includegraphics[width=\linewidth]{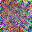}
      \caption*{SSIM = 0.080}
    \end{minipage}
    \caption{\texorpdfstring{$R=0.2, (Acc=0.9568)$}}
    \label{fig_proposed_a}
  \end{subfigure}

  \vspace{1em}

  \begin{subfigure}{\textwidth}
    \centering

    \begin{minipage}{0.2\textwidth}
      \centering
      \includegraphics[width=\linewidth]{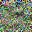}
      \caption*{SSIM = 0.079}
    \end{minipage}
    \begin{minipage}{0.2\textwidth}
      \centering
      \includegraphics[width=\linewidth]{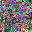}
      \caption*{SSIM = 0.005}
    \end{minipage}
    \begin{minipage}{0.2\textwidth}
      \centering
      \includegraphics[width=\linewidth]{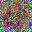}
      \caption*{SSIM = 0.097}
    \end{minipage}
    \begin{minipage}{0.2\textwidth}
      \centering
      \includegraphics[width=\linewidth]{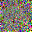}
      \caption*{SSIM = 0.050}
    \end{minipage}

    \caption{\texorpdfstring{$R=0.5, (Acc=0.9525)$}}
    \label{fig_proposed_b}
  \end{subfigure}

  \vspace{1em}

  \begin{subfigure}{\textwidth}
    \centering

    \begin{minipage}{0.2\textwidth}
      \centering
      \includegraphics[width=\linewidth]{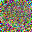}
      \caption*{SSIM = 0.017}
    \end{minipage}
    \begin{minipage}{0.2\textwidth}
      \centering
      \includegraphics[width=\linewidth]{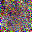}
      \caption*{SSIM = 0.030}
    \end{minipage}
    \begin{minipage}{0.2\textwidth}
      \centering
      \includegraphics[width=\linewidth]{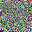}
      \caption*{SSIM = 0.010}
    \end{minipage}
    \begin{minipage}{0.2\textwidth}
      \centering
      \includegraphics[width=\linewidth]{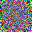}
      \caption*{SSIM = 0.026}
    \end{minipage}

    \caption{\texorpdfstring{$R=0.8, (Acc=0.9536)$}}
    \label{fig_proposed_c}
  \end{subfigure}

  \caption{Images restored by adversarial optimization attacks under FLRSP.}
  \label{fig_proposed_all}
\end{figure}

\begin{figure}[t]
\begin{center}
\includegraphics[width=\linewidth]{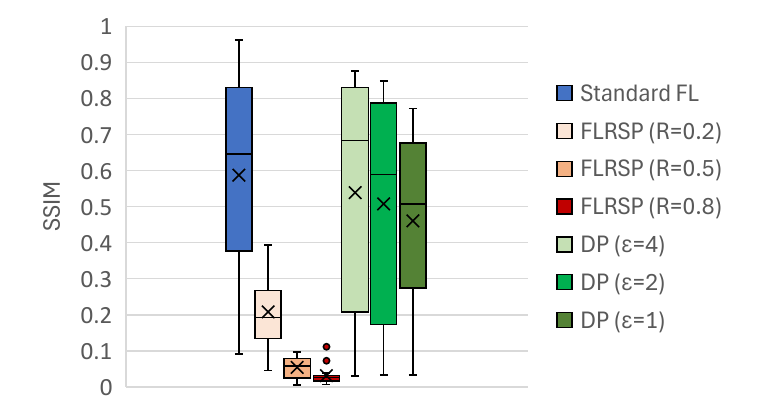}
\caption{Box plots for SSIM comparison of images restored by adversarial optimization attacks under FedSGD (\texorpdfstring{$M=10$}\ \  and \texorpdfstring{$N=5$}\ ). The box represents the interquartile range between the first and third quartiles (\texorpdfstring{$Q1$}\ \  and \texorpdfstring{$Q3$}\ ). The whiskers extend to the minimum and maximum values within the interval \texorpdfstring{$[Q1-1.5(Q3-Q1), Q3+1.5(Q3-Q1)]$}\ . The line inside each box corresponds to the median, while the mean value is indicated by a cross.}
\label{fig8_hokohige}
\end{center}
\end{figure}


Next, we evaluated the robustness of the proposed method against a state-of-the-art attack \cite{cite_AT}, which is an optimization strategy based on adversarial attacks.
Note that the optimization function used for the optimization strategy was Adam, with a learning rate of 0.0001 and 24,000 iterations.
The attack aims to restore images used for FedSGD and requires the gradients and parameters of models to restore images.
In this experiment, the batch size was one and the code for the optimization strategy from \cite{cite_online1} was used. 

Fig.\ref{fig3} shows an example of the attack result and classification accuracy $Acc$ at 10 epochs.
Fig.\ref{fig3}\subref{fig3_a}  shows original training images, and restored images and their SSIM values are given in Fig.\ref{fig3}\subref{fig3_b}.
As shown in the figure, standard FL was not sufficiently robust against the attacks.

Fig.\ref{fig_dp_all} shows images reconstructed from the gradients and parameters of models trained with differential privacy where their original training images were given in Fig.\ref{fig3}\subref{fig3_a}.
As shown in Fig.\ref{fig_dp_all}, the smaller values of $\epsilon$ provided stronger privacy protection.
However, six out of 15 images for every $\epsilon$-value were almost completely restored.
The effectiveness of differential privacy depends on the type of images in general.

Fig.\ref{fig_proposed_all} depicts images restored from the gradients and parameters of the models trained with FLRSP.
As the figure shows, FLRSP was sufficiently resistant to the attack while maintaining almost the same accuracy as that of standard FL.
The proposed method sets a subset of gradients to zero, which disrupts the cosine similarity in Eq.~\eqref{eq_attack} and makes it difficult to generate reconstructed images through iterative image updates. In addition, Fig.\ref{fig8_hokohige} shows the distribution of SSIM values between the original images and the restored ones, which indicated that FLRSP achieved lower SSIM values than DP and standard FL for all $R$ values.
Therefore, FLRSP was demonstrated to maintain high classification accuracy and be more robust against the attack than DP.

\subsection{Discussion}

\begin{figure}[t]
    \centering
        \includegraphics[width=10cm]{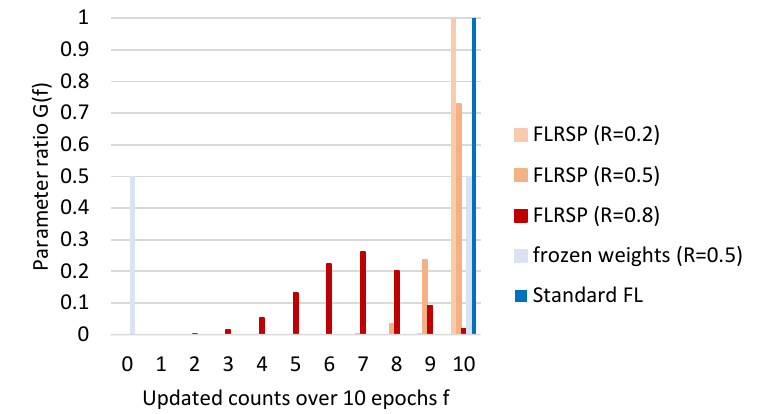}
    \label{4_g2}
\caption{Comparison of parameter ratio under FedAvg (\texorpdfstring{$M=10, N=5$}\ ).}
\label{fig_7}
\end{figure}

Here, we discuss FLRSP in terms of the relationship between classification accuracy $Acc$ and probability $R$ under FedAvg.
If $R=1$ is chosen by each client, all corresponding parameters are not updated. When each client also uses a common random sequence $B^{n, m}_{k, i}$ with $0<R<1$ across all epochs as
\begin{equation}
B^{n, m}_{k, i} = B^{p, m}_{k, i}, (n,p \in \big\{1, 2, ..., N\big\}),
\label{eq:Pf}
\end{equation}
some of the parameters are not updated.
This FL is called FL with frozen weights.
In contrast, in FLRSP, each client uses a different sequence that changes at each epoch.
As a result, the proportion of non-updated parameters is significantly reduced even when using $R<1$. 
Here, let $G(f)$ denote the ratio of parameters updated $f$ times during training over $M$ epochs.
$G(f)$ is given by
\begin{equation}
G(f) = {}_M C_f \times (1-R^{N})^{f} \times R^{N \times (1-f)},
\label{eq:Pf2}
\end{equation}
where $M$ and $N$ denote the number of epochs and the number of clients, respectively.
Here, ${}_M C_f$ represents the number of combinations.
Fig.\ref{fig_7} shows the update ratio of parameters when $M=10$ and $N=5$.

As shown in the figure, FLRSP with $R=0.2$ has almost the same $G(f)$ as that of standard FL. Furthermore, for $R=0.5$, most of the parameters are also updated more than nine times out of ten. In contrast, many parameters used in FL with frozen weights are not updated; when $R=0.5$, half of the parameters were not updated as in Fig.\ref{fig_7}. The fixed-position method \cite{cite_6}, which is a type of FL with frozen weights, cannot update the parameters of the positional embedding layer.
In contrast, for FLRSP, all parameters can be updated by changing random sequences every epoch. In addition, FLRSP exhibited sufficient robustness against both APRIL and the adversarial optimization attacks even with $R=0.2$. FLRSP can also be applied to arbitrary models including ViT and ResNet-based ones, but APRIL is limited to ViT.

\section{Conclusion}
\label{section_5}
In this paper, we proposed FLRSP, a security enhancement method for FL that uses randomly selected model parameters. FLRSP was verified to be effective in terms of the robustness against attacks and classification accuracy in image classification tasks.
The effectiveness of FLRSP was experimentally demonstrated on ViT and ResNet34. FLRSP is also expected to be applicable to almost all types of FL. In addition to image classification tasks, we will examine the effectiveness of the method on other tasks.



\printbibliography

@article{cite_pia,
  title={Overlearning reveals sensitive attributes},
  author={Song, Congzheng and Shmatikov, Vitaly},
  journal={arXiv preprint arXiv:1905.11742},
  year={2019}
}

@article{cite_member,
  title={Membership inference attacks and defenses in federated learning: A survey},
  author={Bai, Li and Hu, Haibo and Ye, Qingqing and Li, Haoyang and Wang, Leixia and Xu, Jianliang},
  journal={ACM Computing Surveys},
  volume={57},
  number={4},
  pages={1--35},
  year={2024},
  publisher={ACM New York, NY}
}

@article{cite_0,
  title={Multi-task distributed learning using vision transformer with random patch permutation},
  author={Park, Sangjoon and Ye, Jong Chul},
  journal={IEEE Transactions on Medical Imaging},
  volume={42},
  number={7},
  pages={2091--2105},
  year={2022},
  publisher={IEEE}
}

@inproceedings{cite_1,
  title={Communication-efficient learning of deep networks from decentralized data},
  author={McMahan, Brendan and Moore, Eider and Ramage, Daniel and Hampson, Seth and y Arcas, Blaise Aguera},
  booktitle={Artificial intelligence and statistics},
  pages={1273--1282},
  year={2017},
  organization={PMLR}
}

@article{cite_2,
  title={Fishing for user data in large-batch federated learning via gradient magnification},
  author={Wen, Yuxin and Geiping, Jonas and Fowl, Liam and Goldblum, Micah and Goldstein, Tom},
  journal={arXiv preprint arXiv:2202.00580},
  year={2022}
}

@article{cite_3,
  title={Deep leakage from gradients},
  author={Zhu, Ligeng and Liu, Zhijian and Han, Song},
  journal={Advances in neural information processing systems},
  volume={32},
  year={2019}
}

@inproceedings{cite_4,
  title={Exploiting unintended feature leakage in collaborative learning},
  author={Melis, Luca and Song, Congzheng and De Cristofaro, Emiliano and Shmatikov, Vitaly},
  booktitle={2019 IEEE symposium on security and privacy (SP)},
  pages={691--706},
  year={2019},
  organization={IEEE}
}

@inproceedings{cite_5,
  title={Gradvit: Gradient inversion of vision transformers},
  author={Hatamizadeh, Ali and Yin, Hongxu and Roth, Holger R and Li, Wenqi and Kautz, Jan and Xu, Daguang and Molchanov, Pavlo},
  booktitle={Proceedings of the IEEE/CVF Conference on Computer Vision and Pattern Recognition},
  pages={10021--10030},
  year={2022}
}

@inproceedings{cite_6,
  title={April: Finding the achilles' heel on privacy for vision transformers},
  author={Lu, Jiahao and Zhang, Xi Sheryl and Zhao, Tianli and He, Xiangyu and Cheng, Jian},
  booktitle={Proceedings of the IEEE/CVF Conference on Computer Vision and Pattern Recognition},
  pages={10051--10060},
  year={2022}
}

@article{cite_AT,
  title={Inverting gradients-how easy is it to break privacy in federated learning?},
  author={Geiping, Jonas and Bauermeister, Hartmut and Dr{\"o}ge, Hannah and Moeller, Michael},
  journal={Advances in neural information processing systems},
  volume={33},
  pages={16937--16947},
  year={2020}
}

@article{cite_11,
  title={Differentially private empirical risk minimization.},
  author={Chaudhuri, Kamalika and Monteleoni, Claire and Sarwate, Anand D},
  journal={Journal of Machine Learning Research},
  volume={12},
  number={3},
  year={2011}
}

@inproceedings{cite_12,
  title={A hybrid approach to privacy-preserving federated learning},
  author={Truex, Stacey and Baracaldo, Nathalie and Anwar, Ali and Steinke, Thomas and Ludwig, Heiko and Zhang, Rui and Zhou, Yi},
  booktitle={Proceedings of the 12th ACM workshop on artificial intelligence and security},
  pages={1--11},
  year={2019}
}

@inproceedings{cite_13,
  title={Federated learning with bayesian differential privacy},
  author={Triastcyn, Aleksei and Faltings, Boi},
  booktitle={2019 IEEE International Conference on Big Data (Big Data)},
  pages={2587--2596},
  year={2019},
  organization={IEEE}
}

@article{cite_14,
  title={Differential privacy for deep and federated learning: A survey},
  author={El Ouadrhiri, Ahmed and Abdelhadi, Ahmed},
  journal={IEEE access},
  volume={10},
  pages={22359--22380},
  year={2022},
  publisher={IEEE}
}

@article{cite_15,
  title={Federated learning with differential privacy: Algorithms and performance analysis},
  author={Wei, Kang and Li, Jun and Ding, Ming and Ma, Chuan and Yang, Howard H and Farokhi, Farhad and Jin, Shi and Quek, Tony QS and Poor, H Vincent},
  journal={IEEE transactions on information forensics and security},
  volume={15},
  pages={3454--3469},
  year={2020},
  publisher={IEEE}
}

@article{cite_17,
  title={An image is worth 16x16 words: Transformers for image recognition at scale},
  author={Dosovitskiy, Alexey and Beyer, Lucas and Kolesnikov, Alexander and Weissenborn, Dirk and Zhai, Xiaohua and Unterthiner, Thomas and Dehghani, Mostafa and Minderer, Matthias and Heigold, Georg and Gelly, Sylvain and others},
  journal={arXiv preprint arXiv:2010.11929},
  year={2020}
}

@INPROCEEDINGS{cite_19,
  author={Sawada, Hiroto and Imaizumi, Shoko and Kiya, Hitoshi},
  booktitle={2024 Asia Pacific Signal and Information Processing Association Annual Summit and Conference (APSIPA ASC)}, 
  title={Enhancing Security Using Random Binary Weights in Privacy-Preserving Federated Learning}, 
  year={2024},
  volume={},
  number={},
  pages={1-6}
}

@inproceedings{cite_20,
  title={Enhanced Security with Encrypted Vision Transformer in Federated Learning},
  author={Aso, Rei and Shiota, Sayaka and Kiya, Hitoshi},
  booktitle={2023 IEEE 12th Global Conference on Consumer Electronics (GCCE)},
  pages={819--822},
  year={2023},
  organization={IEEE}
}

@article{cite_online1,
  author = "Ligeng, Zhu and Luke, Dong",
  title={Deep Leakage From Gradients},
  url={https://github.com/mit-han-lab/dlg},
  chapter =      "",
  pages={14774--14784},
  number =       "",
  type =         "",
  month =        "",
  note =         "",
  year =         "2019"
}

@article{cite_online2,
  author = "Ross, Wightman",
  title={PyTorch Image Models},
  url={https://github.com/huggingface/pytorch-image-models},
  chapter =      "",
  pages =        "",
  number =       "",
  type =         "",
  month =        "",
  note =         "",
  year =         "2025"
}

@inproceedings{cite_addDP_1,
  title={Federated learning of gboard language models with differential privacy},
  author={Xu, Zheng and Zhang, Yanxiang and Andrew, Galen and Choquette, Christopher and Kairouz, Peter and Mcmahan, Brendan and Rosenstock, Jesse and Zhang, Yuanbo},
  booktitle={Proceedings of the 61st Annual Meeting of the Association for Computational Linguistics (Volume 5: Industry Track)},
  pages={629--639},
  year={2023}
}

@article{cite_addDP_2,
  title={A systematic survey for differential privacy techniques in federated learning},
  author={Zhang, Yi and Lu, Yunfan and Liu, Fengxia},
  journal={Journal of Information Security},
  volume={14},
  number={2},
  pages={111--135},
  year={2023},
  publisher={Scientific Research Publishing}
}

@inproceedings{cite_addDP_3,
  title={Agc-dp: Differential privacy with adaptive gaussian clipping for federated learning},
  author={Hidayat, Muhammad Ayat and Nakamura, Yugo and Dawton, Billy and Arakawa, Yutaka},
  booktitle={2023 24th IEEE international conference on mobile data management (MDM)},
  pages={199--208},
  year={2023},
  organization={IEEE}
}

@article{cite_addDP_4,
  title={A federated learning differential privacy algorithm for non-Gaussian heterogeneous data},
  author={Yang, Xinyu and Wu, Weisan},
  journal={Scientific Reports},
  volume={13},
  number={1},
  pages={5819},
  year={2023},
  publisher={Nature Publishing Group UK London}
}

@article{cite_addDP_5,
  title={Federated learning with differential privacy on personal opinions: a privacy-preserving approach},
  author={Ahmadzai, Mirwais and Nguyen, Giang},
  journal={Procedia Computer Science},
  volume={225},
  pages={543--552},
  year={2023},
  publisher={Elsevier}
}

@article{cite_addDP_7,
  title={Federated learning with differential privacy},
  author={Banse, Adrien and Kreischer, Jan and others},
  journal={arXiv preprint arXiv:2402.02230},
  year={2024}
}

@article{cite_addDP_8,
  title={Fedlap-dp: Federated learning by sharing differentially private loss approximations},
  author={Wang, Hui-Po and Chen, Dingfan and Kerkouche, Raouf and Fritz, Mario},
  journal={arXiv preprint arXiv:2302.01068},
  year={2023}
}

@inproceedings{cite_addDP_9,
  title={Towards the robustness of differentially private federated learning},
  author={Qi, Tao and Wang, Huili and Huang, Yongfeng},
  booktitle={Proceedings of the AAAI Conference on Artificial Intelligence},
  volume={38},
  number={18},
  pages={19911--19919},
  year={2024}
}

@article{cite_addDP_10,
  title={Differentially private federated learning with time-adaptive privacy spending},
  author={Kiani, Shahrzad and Kulkarni, Nupur and Dziedzic, Adam and Draper, Stark and Boenisch, Franziska},
  journal={arXiv preprint arXiv:2502.18706},
  year={2025}
}

@article{cite_addAT_11,
  title={Do gradient inversion attacks make federated learning unsafe?},
  author={Hatamizadeh, Ali and Yin, Hongxu and Molchanov, Pavlo and Myronenko, Andriy and Li, Wenqi and Dogra, Prerna and Feng, Andrew and Flores, Mona G and Kautz, Jan and Xu, Daguang and others},
  journal={IEEE Transactions on Medical Imaging},
  volume={42},
  number={7},
  pages={2044--2056},
  year={2023},
  publisher={IEEE}
}

@inproceedings{cite_addAT_12,
  title={Gradient obfuscation gives a false sense of security in federated learning},
  author={Yue, Kai and Jin, Richeng and Wong, Chau-Wai and Baron, Dror and Dai, Huaiyu},
  booktitle={32nd USENIX security symposium (USENIX Security 23)},
  pages={6381--6398},
  year={2023}
}

@inproceedings{cite_addAT_13,
  title={Unveiling Privacy Risks in Stochastic Neural Networks Training: Effective Image Reconstruction from Gradients},
  author={Chen, Yiming and Yang, Xiangyu and Deligiannis, Nikos},
  booktitle={European Conference on Computer Vision},
  pages={397--413},
  year={2024},
  organization={Springer}
}

@article{cite_addAT_14,
  title={Improved gradient leakage attack against compressed gradients in federated learning},
  author={Ding, Xuyang and Liu, Zhengqi and You, Xintong and Li, Xiong and Vasilakos, Athhanasios V},
  journal={Neurocomputing},
  volume={608},
  pages={128349},
  year={2024},
  publisher={Elsevier}
}

@article{cite_addAT_15,
  title={Spear: Exact gradient inversion of batches in federated learning},
  author={Dimitrov, Dimitar I and Baader, Maximilian and M{\"u}ller, Mark and Vechev, Martin},
  journal={Advances in Neural Information Processing Systems},
  volume={37},
  pages={106768--106799},
  year={2024}
}

@inproceedings{cite_APRIL_2,
  title={$\{$SoK$\}$: Gradient Inversion Attacks in Federated Learning},
  author={Carletti, Vincenzo and Foggia, Pasquale and Mazzocca, Carlo and Parrella, Giuseppe and Vento, Mario},
  booktitle={34th USENIX Security Symposium (USENIX Security 25)},
  pages={6439--6459},
  year={2025}
}

@article{cite_APRIL_3,
  title={A Lifecycle-Oriented Survey of Emerging Threats and Vulnerabilities in Large Language Models},
  author={De Maio, Carmen and Di Gisi, Maria and Fenza, Giuseppe and Gallo, Mariacristina and Loia, Vincenzo},
  journal={IEEE Access},
  year={2025},
  publisher={IEEE}
}

@inproceedings{cite_CNNAT_3,
  title={See through gradients: Image batch recovery via gradinversion},
  author={Yin, Hongxu and Mallya, Arun and Vahdat, Arash and Alvarez, Jose M and Kautz, Jan and Molchanov, Pavlo},
  booktitle={Proceedings of the IEEE/CVF conference on computer vision and pattern recognition},
  pages={16337--16346},
  year={2021}
}

@article{cite_kiya_1,
  title={An overview of compressible and learnable image transformation with secret key and its applications},
  author={Kiya, Hitoshi and Maung, April Pyone Maung and Kinoshita, Yuma and Imaizumi, Shoko and Shiota, Sayaka and others},
  journal={APSIPA Transactions on Signal and Information Processing},
  volume={11},
  number={1},
  year={2022},
  publisher={Now Publishers, Inc.}
}

@inproceedings{cite_kiya_2,
  title={Encryption inspired adversarial defense for visual classification},
  author={Maung, Maung and Pyone, April and Kiya, Hitoshi},
  booktitle={2020 IEEE International Conference on Image Processing (ICIP)},
  pages={1681--1685},
  year={2020},
  organization={IEEE}
}

@article{add_DP_125,
  title={The algorithmic foundations of differential privacy},
  author={Dwork, Cynthia and Roth, Aaron and others},
  journal={Foundations and trends{\textregistered} in theoretical computer science},
  volume={9},
  number={3--4},
  pages={211--407},
  year={2014},
  publisher={Now Publishers, Inc.}
}

@article{cite_kiya_3,
  title={Privacy-preserving content-based image retrieval using compressible encrypted images},
  author={Iida, Kenta and Kiya, Hitoshi},
  journal={IEEE Access},
  volume={8},
  pages={200038--200050},
  year={2020},
  publisher={IEEE}
}

@article{cite_kiya_4,
  title={Deepipr: Deep neural network ownership verification with passports},
  author={Fan, Lixin and Ng, Kam Woh and Chan, Chee Seng and Yang, Qiang},
  journal={IEEE Transactions on Pattern Analysis and Machine Intelligence},
  volume={44},
  number={10},
  pages={6122--6139},
  year={2021},
  publisher={IEEE}
}

@article{cite_0224_SSIM,
  title={Image quality assessment: from error visibility to structural similarity},
  author={Wang, Zhou and Bovik, Alan C and Sheikh, Hamid R and Simoncelli, Eero P},
  journal={IEEE transactions on image processing},
  volume={13},
  number={4},
  pages={600--612},
  year={2004},
  publisher={IEEE}
}

@inproceedings{cite_0224_thre,
  title={Refiner: Data refining against gradient leakage attacks in federated learning},
  author={Fan, Mingyuan and Chen, Cen and Wang, Chengyu and Li, Xiaodan and Zhou, Wenmeng},
  booktitle={34th USENIX Security Symposium (USENIX Security 25)},
  pages={3005--3024},
  year={2025}
}

@inproceedings{cite_0224_syusoku,
  title={Energy efficient federated learning with age-weighted FedSGD},
  author={Wang, Kaidi and Ding, Zhiguo and So, Daniel KC and Ding, Zhi},
  booktitle={2024 IEEE International Conference on Communications Workshops (ICC Workshops)},
  pages={457--462},
  year={2024},
  organization={IEEE}
}

@article{0224_AT_Avg,
  title={Data leakage in federated averaging},
  author={Dimitrov, Dimitar Iliev and Balunovic, Mislav and Konstantinov, Nikola and Vechev, Martin},
  journal={Transactions on Machine Learning Research},
  year={2022}
}
%

\end{document}